\documentclass[12pt,epsfig]{article}
\usepackage{epsfig}
\usepackage{amssymb}
\newcommand{\singlespace}{
     \renewcommand{\baselinestretch}{1}\large\normalsize}
\newcommand{\doublespace}{
     \renewcommand{\baselinestretch}{1.6}\large\normalsize}

\newcommand{\beq}{\begin{equation}}
\newcommand{\eeq}{\end{equation}}
\newcommand{\bea}{\begin{eqnarray}}
\newcommand{\eea}{\end{eqnarray}}

\newcommand{\ave}[1]{\langle {#1} \rangle}

\newcommand{\pslash}{p\!\!\!/}

\newcommand{\pb}{\bar\psi}
\newcommand{\qq}{\ave{\pb\psi}}
\newcommand{\dg}{\delta g_{\pi qq}^{-2}}

\def\roughly#1{\mathrel{\raise.3ex\hbox{$#1$\kern-.75em%
\lower1ex\hbox{$\sim$}}}}
\def\lsim{\roughly<}

\def\psl{p\hspace{-1.7mm}/}

\def\dfp{\frac{d^4 p}{(2\pi)^4}}
\def\dfk{\frac{d^4 k}{(2\pi)^4}}
\def\intp{\int\dfp}
\def\intk{\int\dfk}
\def\rrr{\longrightarrow}
\def\={\;=\;}
\def\+{\;+\;}
\def\Tr{{\rm Tr}}
\def\eps{\varepsilon}
\begin{document}
\begin{flushright}
January 2000
\end{flushright}
\vspace{1.0cm}
\begin{center}
\doublespace
\begin{large}
{\bf Meson Properties in the $1/N_c$-corrected NJL model}\\
\end{large}
\vskip 1.0in
M. Oertel, M. Buballa and J. Wambach\\
{\small{\it Institut f\"ur Kernphysik, TU Darmstadt,\\ 
Schlossgartenstr. 9, 64289 Darmstadt, Germany}}\\
\end{center}
\vspace{1cm}

\begin{abstract}
Properties of mesons are investigated within the Nambu--Jona-Lasinio 
model. We include meson-loop corrections, which are generated via a 
systematic $1/N_c$-expansion in next-to-leading order.
We show that our scheme is consistent with chiral symmetry, in particular
with the Goldstone theorem and the Gell-Mann Oakes Renner relation.
The numerical part focuses on the pion and the $\rho$-meson
sector. For the latter the $1/N_c$-corrections are crucial in order
to include the dominant $\rho \rightarrow \pi\pi$-decay channel,
while the leading-order approximation only contains unphysical 
$q\bar q$-decay channels. We show that a satisfactory description of 
the pion electromagnetic form factor can be obtained. Similarities
and differences to hadronic models are discussed.

\end{abstract}

\newpage

\singlespace

\section{Introduction}

The understanding of hadron properties in the vacuum as well as in hot or 
dense matter is one of the central tasks of present-day nuclear physics. 
In principle, all properties of strongly interacting particles
should be derived from QCD. 
However, at least in the low-energy regime, where perturbation theory 
is not applicable, this is presently limited to a rather small number of 
observables which can be studied on the lattice, while more complex 
processes have to be described within model calculations.

So far the best descriptions of hadronic spectra, decays and
scattering processes are obtained within phenomenological hadronic models.
For instance the pion electromagnetic form factor in the time-like
region can be reproduced rather well within a simple vector dominance
model with a dressed $\rho$-meson which is constructed by coupling a
bare $\rho$-meson to a two-pion intermediate state \cite{brown,herrmann}. 
Models of this type have been successfully extended to investigate medium 
modifications of vector mesons and to calculate dilepton production rates
in hot and dense hadronic matter \cite{rapp}.     

In this situation one might ask how the phenomenologially successful
hadronic models emerge from the underlying quark structure and the
symmetry properties of QCD. Since this question cannot be answered 
at present from first principles it has to be addressed within quark 
models. For light hadrons chiral symmetry and its spontaneous breaking in 
the physical vacuum through instantons plays the decisive role in 
describing the two-point correlators \cite{schaefer} with 
confinement being much less important. This feature is 
captured by the Nambu--Jona-Lasinio(NJL) model in which the four-fermion
interactions can be viewed as being induced by instantons.

The study of hadrons within the NJL model has of course a long history.
In fact, mesons of various quantum numbers have already been discussed
in the original papers by Nambu and Jona-Lasinio \cite{njl} and by
many authors thereafter (for reviews see \cite{vogl,klevansky,hatsuda}).
Most of these works correspond to a leading-order approximation in
$1/N_c$, the inverse number of colors. In this scheme quark masses are
calculated in mean-field approximation and 
mesons are constructed as correlated quark-antiquark states.
With the appropriate choice of parameters chiral symmetry, which is
an (approximate) symmetry of the model Lagrangian,
is spontaneously broken in the vacuum and pions emerge as (nearly) massless 
Goldstone bosons. While this is clearly one of the successes of the model, 
the description of other mesons is more problematic. 
One reason is the fact that the NJL model does not confine quarks.
As a consequence a meson can decay into free constituent quarks if its 
mass is larger
than twice the constituent quark mass $m$. Hence, for a typical value of
$m \sim$~300~MeV, the $\rho$-meson with a mass of 770~MeV, for
instance, would be unstable against decay into quarks.   
On the other hand the physical decay channel of the $\rho$-meson into two
pions is not included in the standard approximation. 
Hence, even if a large constituent quark mass is chosen in order to suppress 
the unphysical decays into quarks, one obtains a poor description of the 
$\rho$-meson propagator and related observables, like the pion electromagnetic 
form factor.

This and other reasons have motivated several authors to go beyond the 
standard approximation scheme and to include mesonic fluctuations.
In Ref.~\cite{krewald} a quark-antiquark $\rho$-meson is coupled via a quark
triangle to a two-pion state. 
Also higher-order corrections to the quark self-energy \cite{quack}
and to the quark condensate \cite{blaschke} have been investigated.
However, as the most important feature of the NJL model is chiral symmetry,
one should use an approximation scheme which conserves the symmetry 
properties, to ensure the existence of massless Goldstone bosons.

Two slightly different symmetry conserving approximation schemes are 
discussed in Refs.~\cite{dmitrasinovic} and \cite{nikolov}. 
The authors of Ref.~\cite{nikolov} use an expansion in $1/N_c$ in order to
include mesonic fluctuations and calculate the changes of $f_{\pi}$ and of 
the quark condensate $\ave{\bar\psi\psi}$ in a low-momentum expansion for the 
incoming state.
In Ref.~\cite{dmitrasinovic} a correction term to the quark self-energy
is included in the gap equation. In a perturbative scheme this term 
would be of order $1/N_c$ but, as the modified gap equation is solved
self-consistently, arbitrary orders in $1/N_c$ are generated.
The authors find a consistent scheme to describe mesons and show the
validity of the Goldstone theorem and the Goldberger-Treiman relation
in that scheme. 
Based on this model various authors have investigated the effect of
meson-loop corrections on the pion electromagnetic form factor
\cite{lemmer} and on $\pi$-$\pi$ scattering in the vector \cite{he}
and the scalar channel \cite{huang}. 
However, since the numerical calculation of the multi-loop diagrams,
which have to be evaluated, is rather involved, in these references
the exact expressions are approximated by low-momentum expansions. 

In the present paper we calculate the $\rho$-meson self energy
in a strict expansion up to next-to-leading order without any approximations.
Within the same scheme we have recently studied the influence of mesonic 
fluctuations on the pion propagator \cite{oertel}. This was mainly motivated
by recent works by Kleinert and Van den Bossche \cite{kleinert}, who claim 
that chiral symmetry is {\it not} spontaneously broken in the NJL model as 
a result of strong mesonic fluctuations. In Ref.~\cite{oertel} we argue 
that because of the non-renormalizability of the NJL model new divergences 
and hence new cutoff parameters emerge if one includes meson loops.
Following Refs.~\cite{dmitrasinovic} and \cite{nikolov} we regularize the
meson loops by an independent cutoff parameter $\Lambda_M$. The results 
are, of course, strongly dependent on this parameter. Whereas for moderate 
values of $\Lambda_M$ the pion properties change only quantitatively 
strong instabilities are encountered for larger values of $\Lambda_M$, which might 
be a hint for an instability of the spontaneously broken vacuum state. 
 
In Ref.~\cite{oertel} we restricted ourselves to calculate the pion mass
$m_\pi$, the pion decay constant $f_\pi$ and the quark condensate
$\ave{\bar\psi\psi}$. Since these quantities can already be fitted without
meson-loop corrections (i.e. $\Lambda_M$~=~0) one has to look at other
observables in order to fix $\Lambda_M$. Obviously the spectral function
of the $\rho$-meson is particularly suited, as it cannot be described 
realistically without taking into account pion loops.
In the present article all parameters of the NJL model will be fixed by
fitting $\rho$-meson properties simultaneously with $f_\pi$ and 
$\ave{\bar\psi\psi}$. The important result is that such a fit can indeed be
achieved with a constituent quark mass which is large enough such that the 
unphysical $q\bar q$-threshold opens above the $\rho$-meson peak. Since
the constituent quark mass is not an independent input parameter this
was not clear a priori. 

The paper is organized as follows.
In Sec.~\ref{model} we present the scheme for describing mesons
in next-to-leading order in $1/N_c$. The consistency of this scheme
with the Goldstone theorem and with the Gell-Mann Oakes Renner relation
will be shown in Sec.~\ref{pion}. In Sec.~\ref{hadron} we 
perform a low-momentum expansion of the effective meson vertices
and discuss the relation to hadronic models.
The numerical results will be presented in Sec.~\ref{numerics}.
For several values of $\Lambda_M$ we fix a subset of the model parameters
by fitting quantities in the pion sector. Then we determine the
remaining parameters, in particular $\Lambda_M$, in the $\rho$-meson sector.
Finally, conclusions are drawn in Sec.~\ref{conclusions}.

\section{The NJL model in leading order and next-to-leading order in $1/N_c$}
\label{model}

We consider the following NJL-type Lagrangian:
\beq
   {\cal L} \;=\; \pb ( i \partial{\hskip-2.0mm}/ - m_0) \psi
            \;+\; g_s\,[(\pb\psi)^2 + (\pb i\gamma_5{\vec\tau}\psi)^2]  
            \;-\; g_v\,[(\pb\gamma^\mu{\vec\tau}\psi)^2 + 
                        (\pb\gamma^\mu\gamma_5{\vec\tau}\psi)^2]   
   \,.
\eeq
Here $\psi$ is a quark field with $N_f$~=~2 flavors and $N_c$ colors,
while $g_s$ and $g_v$ are dimensionful coupling constants of the order
$1/N_c$. In order to establish the counting scheme, the number of colors
has been treated as variable, but all numerical calculations will be done
with the physical value, $N_c$~=~3.

\begin{figure}[b!]
\hspace{3cm}
\parbox{10cm}{
     \epsfig{file=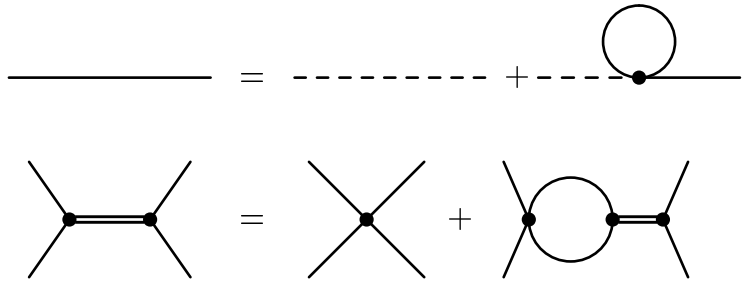,
     height=3.7cm, width=10.0cm}}
\caption{\it (Upper part) The Dyson equation for the quark propagator in the 
         Hartree approximation (solid lines). The dashed lines denote
         the bare quark propagator.  
         (Lower part) The Bethe-Salpeter equation for the meson propagator 
         in the RPA (double line). Again, the single solid lines indicate
         quark propagators in the Hartree approximation.}
\label{fig1} 
\end{figure}

In most publications the NJL model has been treated in leading order
in $1/N_c$. In terms of many-body theory this corresponds to a 
(Bogoliubov) Hartree
approximation for the quark propagator and to a random phase approximation
(RPA) for describing mesonic excitations. Diagrammatically this is shown in 
Fig.~\ref{fig1}. The selfconsistent solution of the Dyson equation shown 
in the upper part of Fig.~\ref{fig1} leads to a momentum independent quark 
self energy and therefore only gives a correction to the constituent quark mass:
\beq
m \= m_0 \+ 2i g_s\ 4 N_c N_f \intp {m\over{p^2-m^2+i\epsilon}}~.
\label{gap}
\eeq
Usually one refers to this equation as the gap equation. 
For sufficiently large couplings $g_s$ it allows for a finite constituent
quark mass $m$ even in the chiral limit, i.e. for $m_0$~=~0. In the 
mean-field approximation this solution minimizes the ground-state energy.
Since $g_s$ is of order $1/N_c$ the constituent quark mass $m$, and hence 
the quark propagator $S(p) = (\psl - m)^{-1}$ are of the order unity. 

Mesons are described via a Bethe-Salpeter equation, as shown in the
lower part of Fig.~\ref{fig1}. Since this is all standard we only list 
the results, which are needed later on.
First we define the quark-antiquark polarization functions
\beq
   \Pi_M(q)\= -i\intp \Tr[\,\Gamma_M \, iS(p+{q\over2})
                  \,\Gamma_M \, iS(p-{q\over2})\,] \;,
\label{pol0}
\eeq
with $M = \sigma, \pi, \rho, a_1$ and
$\Gamma_\sigma = 1\!\!1$, $\Gamma_\pi^k = i\gamma_5\tau^k$,
$\Gamma_\rho^{\mu\,k} = \gamma^\mu\tau^k$ and 
$\Gamma_{a_1}^{\mu\,k} = \gamma^\mu\gamma_5\tau^k$.
Here ``$\Tr$'' denotes a trace in color, flavor and Dirac space.
$\Pi_M$ is diagrammatically shown in Fig.~\ref{fig3}.
Iterating the scalar (pseudoscalar) part of the four-fermion 
interaction one obtains for the sigma meson (pion):
\beq
D_\sigma(q) \= \frac{-2 g_s}{1-2g_s\Pi_\sigma(q)} \;,\qquad
D_\pi^{ab}(q) \;\equiv\;D_\pi(q)\,\delta_{ab} 
\= \frac{-2 g_s}{1-2g_s\Pi_\pi(q)}\,\delta_{ab}\;.
\label{dsigmapi}
\eeq
Here $a$ and $b$ are isospin indices and we have used the notation
$\Pi_\pi^{ab}(q) \equiv \Pi_\pi(q)\,\delta_{ab}$.

In the vector channel this can be done in a similar way. Using the
transverse structure of the polarization loop in the vector channel,
\beq
    \Pi_\rho^{\mu\nu, ab}(q) \= \Pi_\rho(q)\,T^{\mu\nu}\, \delta_{ab}
    \;;\qquad T^{\mu\nu} =  (-g^{\mu\nu} + \frac{q^\mu q^\nu}{q^2})
    \;,
\label{pirho}
\eeq
one obtains for the $\rho$-meson
\beq
D_\rho^{\mu\nu,ab}(q) \;\equiv\; D_\rho(q)\;T^{\mu\nu}\;\delta_{ab}
\= \frac{-2 g_v}{1-2g_v\Pi_\rho(q)} \,T^{\mu\nu}\;\delta_{ab} 
 \;.
\label{drho}
\eeq
Analogously, the $a_1$ can be constructed from the transverse part of 
the axial polarization function $\Pi_{a_1}$.
As discussed e.g. in Ref.~\cite{klimt} $\Pi_{a_1}^{\mu\nu}$ also contains 
a longitudinal part which contributes to the pion. Although there is no 
conceptional problem to include this mixing we will neglect it in the 
present paper in order to keep the structure of the model as simple as 
possible.

It follows from Eqs.~(\ref{pol0})~-~(\ref{drho}) that the functions
$D_M(q)$ are of order $1/N_c$. Their explicit forms are given
in App.~\ref{correlators}. For simplicity we will call them
``propagators'', although strictly speaking, they should be interpreted as 
the product of a renormalized meson propagator with a squared 
quark-meson coupling constant.
The latter is given by the inverse residue of the function $D_M(q)$,
while the pole position determines the meson mass:
\beq
   D_M^{-1}(q)|_{q^2 = m_M^{2 (0)}} \= 0 \;,\qquad
    g_{Mqq}^{-2 (0)} \= \frac{d\Pi_M(q)}{dq^2}|_{q^2 = m_M^{2 (0)}} \;.
\label{mesonmass0}
\eeq
We have used the superscript $(0)$ to indicate that $m_M^{2 (0)}$
and $g_{Mqq}^{-2 (0)}$ are leading-order quantities in $1/N_c$. 
One easily verifies that they are of order unity and $1/\sqrt{N_c}$,
respectively.
With the help of the gap equation, Eq.~(\ref{gap}), one can show
that the pion is massless in the chiral limit, demonstrating
the consistency of the scheme with chiral symmetry \cite{njl}.
\begin{figure}[b!]
\hspace{5cm}
\parbox{6cm}{
     \epsfig{file=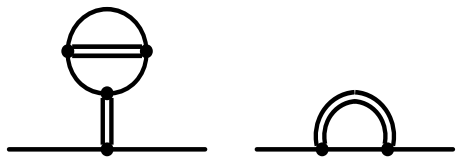,
     height=2.cm, width=6.0cm}}
\caption{\it The $1/N_c$-corrections to the quark self-energy.}
\label{fig2} 
\end{figure}

We now turn to the corrections in next-to-leading order in $1/N_c$.
The correction terms to the quark self-energy are shown in Fig.~\ref{fig2}. 
In these diagrams the quark lines and the double lines
correspond to quark propagators in the Hartree approximation (order unity)
and to meson propagators in the RPA (order $1/N_c$), respectively. 
Both diagrams are therefore of order $1/N_c$. 

The $1/N_c$-corrected mesonic polarization diagrams read
\beq
    {\tilde \Pi}_M(q) \= \Pi_M(q) \+ \sum_{k=a,b,c,d}\; \delta \Pi_M^{(k)}(q)
\;.
\label{pol1}
\eeq
The four correction terms $\delta \Pi_M^{(a)}$ - $\delta \Pi_M^{(d)}$
together with the leading-order term $\Pi_M$ are shown in Fig.~\ref{fig3}. 
Again the lines in this figure correspond to Hartree quarks and RPA mesons. 
Besides the RPA meson propagators the main building blocks are quark
triangle and box diagrams, which are shown in Fig.~\ref{fig4}.
The triangle diagrams entering into $\delta \Pi_M^{(a)}$ and 
$\delta \Pi_M^{(d)}$ can be interpreted as effective three-meson vertices.
For external mesons $M_1$, $M_2$ and $M_3$ they are given by
\bea
 -i \Gamma_{M_1,M_2,M_3}(q,p) &=& - \intk
  \Big\{ \Tr [\Gamma_{M_1}i S(k)\Gamma_{M_2} i S(k-p)\Gamma_{M_3}
 i  S(k+q)] \nonumber \\
& & \hspace{1.8cm}+ \Tr[\Gamma_{M_1}i S(k-q)\Gamma_{M_3}i S(k+p)\Gamma_{M_2}i
  S(k)]\Big\}~, 
\label{trianglevertex}
\eea
with the operators $\Gamma_M$ as defined below Eq.~(\ref{pol0}).
We have summed over both possible orientations of the quark loop.
\begin{figure}[t!]
\hspace{1cm}
\parbox{14cm}{
     \epsfig{file=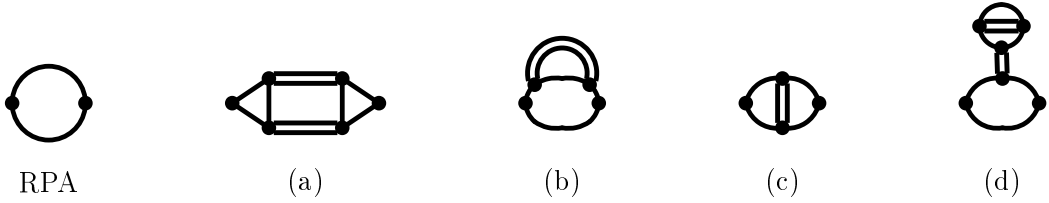,
     height=2.6cm, width=14.0cm}}
\caption{\it Contributions to the mesonic polarization function
             in leading (RPA) and next-to-leading order in $1/N_c$.}
\label{fig3} 
\end{figure}

The quark box diagrams are effective four-meson vertices and are needed 
for the evaluation of $\delta \Pi_M^{(b)}$ and $\delta \Pi_M^{(c)}$.
If one again sums over both orientations of the quark loop they are given
by  
\bea
&&  \hspace{-2.0cm}
-i\Gamma_{M_1,M_2,M_3,M_4}(p_1,p_2,p_3) \phantom{\intk}\nonumber\\
&=&\intk \Big(
\Tr[\Gamma_{M_1}iS(k)\Gamma_{M_2}iS(k-p_2)\Gamma_{M_3}iS(k-p_2-p_3)
    \Gamma_{M_4}iS(k+p_1)]  
\nonumber\\ && \hspace{1.3cm}
 +\Tr[\Gamma_{M_1}iS(k-p_1)\Gamma_{M_4}iS(k+p_2+p_3)\Gamma_{M_3}iS(k+p_2)
      \Gamma_{M_2}iS(k)]\Big)~.
\label{boxvertex}
\eea
\begin{figure}[t]
\hspace{3cm}
\parbox{10cm}{
     \epsfig{file=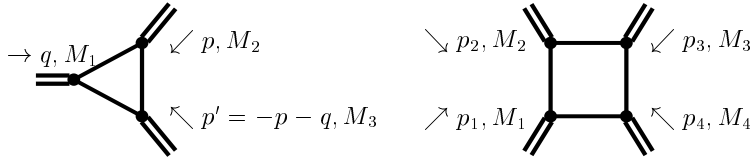,
     height=2.1cm, width=10.cm}}
\caption{\it (Left) The quark triangle vertex $ -i\Gamma_{M_1,M_2,M_3}(q,p)$.
             (Right) The quark box vertex
              $-i\Gamma_{M_1,M_2,M_3,M_4}(p_1,p_2,p_3)$.}
\label{fig4} 
\end{figure}

The polarization diagrams $\delta \Pi_M^{(a)}$ - $\delta \Pi_M^{(d)}$
contain several loops. However, with the help of the definitions 
given above they can be written in a relatively compact form:
\bea
\delta\Pi^{(a)}_{M}(q) &\=& \phantom{-} \frac{i}{2}\intp \sum_{M_1 M_2}
\Gamma_{M,M_1,M_2}(q,p)\, D_{M_1}(p)\,\Gamma_{M,M_1,M_2}(-q,-p)\,
D_{M_2}(-p-q)
\;, \nonumber\\
\delta\Pi^{(b)}_{M}(q)&\=&- i\,\intp \hspace{3.0mm} \sum_{M_1} \;
\Gamma_{M,M_1,M_1,M}(q,p,-p)\,D_{M_1}(p)
\;, \nonumber\\
\delta\Pi^{(c)}_M(q)&\=& -\frac{i}{2}\intp \hspace{2.5mm} \sum_{M_1} \; 
\Gamma_{M,M_1,M,M_1}(q,p,-q)\,D_{M_1}(p)
\;, \nonumber\\
\delta\Pi^{(d)}_M(q) &\=& \frac{i}{2}\;\Gamma_{M,M,\sigma}(q,-q)\, D_\sigma(0)
\; \intp \hspace{3.0mm} \sum_{M_1}\; 
\Gamma_{M_1,M_1,\sigma}(p,-p) \, D_{M_1}(p)
\;.  
\label{deltapi}
\eea
Note that for $\delta\Pi^{(c)}_M$ and $\delta\Pi^{(d)}_M$ one has to include 
a symmetry factor of $1/2$, because otherwise the sum over the two 
orientations of the quark propagators, which is contained in the definitions 
of the quark triangle vertex (Eq.~(\ref{trianglevertex})) and of the quark 
box vertex (Eq.~(\ref{boxvertex})) would lead to double counting.
Similarly in $\delta\Pi^{(a)}_M$ we had to correct for the fact that the
exchange of $M_1$ and $M_2$ leads to identical diagrams.

For the evaluation of Eq.~(\ref{deltapi}) we have to proceed in two steps. 
In the first step we calculate the intermediate RPA meson-propagators.
Simultaneously we can calculate the quark triangles and box diagrams.
One is then left with a meson loop which has to be evaluated in a second step. 

The various sums are, in principle, over all quantum numbers
of the intermediate mesons. Of course, many combinations vanish,
e.g. because of isospin conservation. 
In fact, the expression for $\delta\Pi^{(d)}_M(q)$ should contain a sum over 
the quantum numbers of both intermediate mesons. 
However, one can easily verify that the meson which connects the two
quark triangles has to be a sigma meson in order to give a non-vanishing
contribution.
 
As we have mentioned earlier, the main focus of the present paper is on the 
$\rho$-meson.
In this case the most important contribution comes from the two-pion
intermediate state in diagram $\delta\Pi^{(a)}_M$. 
Other contributions to this diagram, i.e $\pi a_1$, $\rho\sigma$, $\rho\rho$ 
and $a_1 a_1$ intermediate states, are much less important since the
corresponding decay channels open far above the $\rho$-meson mass and
- in the NJL model - also above the unphysical two-quark threshold.
Hence, from a purely phenomenological point of view, it should be sufficient 
to restrict the sums in Eq.~(\ref{deltapi}) to intermediate pions.
However, in order to stay consistent with chiral symmetry, we have to 
include intermediate sigma mesons as well. On the other hand,  
vector- and axial-vector mesons can be neglected without violating chiral symmetry.
Since this leads to an appreciable simplification of the numerics
we have restricted the {\it intermediate} degrees of freedom to 
scalar and pseudoscalar mesons in the present paper.   
Of course, in order to describe a $\rho$-meson, we have to take vector
couplings at the external vertices of the diagrams shown in Fig.~\ref{fig3}. 

In analogy to Eqs.~(\ref{dsigmapi}) and (\ref{drho}) the corrected 
meson propagators are given by
\beq
{\tilde D}_M(q) \= \frac{-2 g_M}{1-2g_M {\tilde\Pi}_{M}(q)} \;,
\label{dmtilde}
\eeq
with $g_M = g_s$ for $\sigma$ and $\pi$ and $g_M = g_v$ for $\rho$ and $a_1$.
Since our scheme corresponds to a $1/N_c$-expansion of the polarization
diagrams and hence of the {\it inverse} meson propagators, the
${\tilde D}_M$ contains terms of arbitrary orders in $1/N_c$.

The corrected meson masses are again defined by the pole positions 
of the propagators:
\beq
   \tilde{D}_M^{-1}(q)|_{q^2 = m_M^2} \= 0 \;.
\label{mesonmass1}
\eeq
As we will show in the next section our scheme is consistent with the
Goldstone theorem, i.e. in the chiral limit it leads to massless pions.
Note, however, that because of its implicit definition $m_M$ contains terms 
of arbitrary orders in $1/N_c$, although we start from a strict expansion of
the inverse meson propagator up to next-to-leading order.
This will be important in the context of the Gell-Mann Oakes Renner relation.

Finally we should comment on Ref.~\cite{dmitrasinovic}, where the authors
introduce $1/N_c$-corrections in a slightly different way. 
In that scheme a correction term to the quark self-energy is 
self-consistently included in the gap equation. 
This leads to a modified quark mass as compared to the 'Hartree mass'. 
Mesons are described by iterating the polarization diagrams 
$\Pi_M$, $\delta\Pi_M^{(a)}$, $\delta\Pi_M^{(b)}$ and $\delta\Pi_M^{(c)}$, 
but using the modified quarks for constructing RPA mesons as
well as quark triangle- and box diagrams.
Since the $1/N_c$-correction terms are iterated in the gap equation, all
diagrams contain terms of arbitrary orders in $1/N_c$. 
However, if one performs a strict $1/N_c$-expansion of the mesonic 
polarization diagrams one exactly recovers our scheme of describing 
mesons~\cite{dmitrasinovic}. 

Both schemes, the one introduced in Ref.~\cite{dmitrasinovic}
and the strict $1/N_c$-expansion, are consistent with chiral symmetry.
In particular, they lead to massless pions in the chiral limit. 
However, because of the modified gap equation, in the scheme of 
Ref.~\cite{dmitrasinovic} the RPA pions are not massless in the chiral 
limit, but tachyonic. Therefore this model is not very well suited
for calculating the $\rho$-meson self-energy, which is mainly determined by 
intermediate RPA pions. For this reason we prefer the 
strict $1/N_c$-expansion scheme, in which both RPA- and  
$1/N_c$-corrected pions are massless in the chiral limit. 
Still there remains the problem that $m_\pi \neq m_\pi^{(0)}$,  if we
go away from the chiral limit, but at least for our final parameter set
the difference is only about 10\%. This will be discussed in more detail
in Sec.~\ref{numerics}.

\section{Consistency with chiral symmetry}
\label{pion}

Before discussing the numerical details we wish to show that 
the scheme introduced in the previous section is consistent with chiral 
symmetry. We begin with the Goldstone theorem and then show the consistency 
of the scheme with the Gell-Mann Oakes Renner relation.
This is not an entirely academic exercise. Since most of the integrals
which have to be evaluated are divergent and hence must be regularized
one has to ensure that the various symmetry relations are not destroyed
by the regularization. To this end, it is important to know how these
relations formally emerge.

The validity of the Goldstone theorem in our scheme has already been proven 
by Dmitra\v{s}inovi\'{c} et al.~\cite{dmitrasinovic} and we only summarize 
the main steps which are relevant for our later discussion.
One has to show that, in the chiral limit, the inverse pion propagator vanishes 
at zero momentum, 
\beq 
    2g_s\,{\tilde\Pi}_\pi(0) \= 1 \qquad {\rm for} \quad m_0 \= 0.
    \label{goldstone}
\eeq
As before we use the notation
${\tilde\Pi}^{ab}_\pi = \delta_{ab}{\tilde\Pi}_\pi$.  
Restricting the calculation to the chiral limit and to zero momentum
simplifies the expressions considerably and Eq.~(\ref{goldstone}) can 
be proven analytically.
Since the Goldstone theorem is fulfilled in leading
order, i.e. $2g_s\,\Pi_\pi(0)$~=~1 for $m_0$~=~0, we only need to show
that the contributions of the correction terms add up to zero:
\beq 
    \sum_{k=a,b,c,d}\;\delta\Pi_\pi^{(k)}(0) \= 0
    \qquad {\rm for} \quad m_0 \= 0.
    \label{deltapisum}
\eeq
Let us begin with diagram $\delta\Pi_\pi^{(a)}$. As mentioned above, 
we neglect the $\rho$ and $a_1$ subspace for intermediate mesons.
Then one can easily see that the external pion can only couple to
a $\pi\sigma$ intermediate state. Evaluating the trace in 
Eq.~(\ref{trianglevertex}) for zero external momentum one gets for
the corresponding triangle diagram:
\beq
  \Gamma_{\pi,\pi,\sigma}^{ab}(0,p) = -\delta_{ab}\ 4  N_c N_f\ 2 m\ I(p)~, 
  \label{gammapps}
\eeq
with $a$ and $b$ being isospin indices and the elementary integral
\beq
    I(p) = \intk \frac{1}{(k^2-m^2+i \eps)( (k+p)^2-m^2+i\eps)}\;.
\eeq
Inserting this into Eq.~(\ref{deltapi}) we find 
\beq
\delta\Pi_\pi^{(a)\,ab}(0) 
\= i \delta_{ab}\intp  (4 N_c N_f I(p))^2 4 m^2\ D_\sigma(p)\ D_\pi(p) \;.
\label{pisigtri}
\eeq
Now the essential step is to realize that the product of the RPA sigma- 
and pion propagators can be converted into a difference \cite{dmitrasinovic}, 
\beq
D_\sigma(p)\ D_\pi(p) =
  i \,\frac{D_\sigma(p)-D_\pi(p)} {4 N_c N_f\ 2 m^2\ I(p)}\;,
\label{pisig}
\eeq
to finally obtain
\bea
\delta\Pi_\pi^{(a)\,ab}(0) =-\delta_{ab}\;4 N_c N_f\intp
2 I(p)\Big\{ D_\sigma(p)-D_\pi(p)\Big\} \;. 
\label{pisigend}
\eea 
The next two diagrams can be evaluated straightforwardly.
One finds:
\bea
  \delta\Pi_\pi^{(b)\,ab}(0)&=&  -\delta_{ab}\;4 N_c N_f\intp\Big\{ 
  D_\sigma(p)\  \big(I(p)+I(0)-(p^2-4 m^2)\ K(p)\big) \nonumber \\
&&\hspace{3.5cm} +D_\pi(p)\ \big(3I(p)\hspace{0.2cm}+\;3I(0)\hspace{0.4cm}
-\;\;3p^2\ K(p)\big)\; \Big\} \;, 
\nonumber\\ 
  \delta\Pi_\pi^{(c)\,ab}(0)&=& -\delta_{ab} \;4 N_c N_f\intp I(p)\Big\{ 
  -D_\sigma(p) - D_\pi(p) \Big\} \;.
\label{pseudo} 
\eea
The elementary integral $K(p)$ is of the same type as the integral
$I(p)$ and is defined in App.~\ref{integrals}.
Finally we have to calculate $\delta\Pi_\pi^{(d)}(0)$.
According to Eq.~(\ref{deltapi}),  it can be written in the form 
\beq
    \delta\Pi_\pi^{(d)\,ab}(0) \= -i\Gamma^{ab}_{\pi,\pi,\sigma}(0,0) \,
    D_\sigma(0)\,\Delta  \;,
    \label{deltapid}
\eeq
with 
\bea
 \Delta &=& \frac{1}{2}\intp\sum_M (-iD_M(p))
 (-i\Gamma_{M,M,\sigma}(p,-p)) 
   \nonumber \\
&=& 4 N_c N_f\ m \int\frac{d^4p}{(2\pi)^4}{\Big\{}
\ D_\sigma(p)\ (2\ I(p)+I(0)-(p^2-4 m^2)\ K(p)) \nonumber \\
&&\hspace{3.3cm} +  D_\pi(p)\  (\;3I(0)\;-\;3p^2\ K(p)\;) 
\hspace{2.5cm}{\Big\}}~. 
\label{tdself}
\eea
Evaluating $D_\sigma(0)$ in the chiral limit and comparing the result with
Eq.~(\ref{gammapps}) one finds that the product of the first two factors 
in Eq.~(\ref{deltapid}) is simply $\delta_{ab}/m$, i.e. one gets   
\beq
 \delta\Pi_\pi^{(d)\,ab}(0) = \delta_{ab} \,\frac{\Delta}{m}~.
\label{pid}
\eeq
With these results one can easily check that Eq.~(\ref{deltapisum}) 
indeed holds in our scheme. 

Of course, most of the integrals we have to deal with are divergent
and have to be regularized. Therefore one has to make sure that all
steps which lead to Eq.~(\ref{deltapisum}) remain valid in the 
regularized model.   
One important observation is that the cancellations occur already
on the level of the $p$-integrand, i.e. before performing the 
meson-loop integral. This means that there is no restriction on the
regularization of this loop. 
We also do not need to perform the various quark loop integrals explicitly
but can make use of several relations between them. For instance,
in order to arrive at Eq.~(\ref{pid}) we need the similar structure
of quark triangle $\Gamma_{\pi,\pi,\sigma}(0,0)$ and the inverse RPA 
propagator $D_\sigma(0)^{-1}$. Therefore all quark loops, i.e. RPA 
polarizations, 
triangles and box diagrams should be consistently regularized within
the same scheme, whereas the meson loops can be regularized independently.

Going away from the chiral limit the pion gets a finite mass.
To lowest order in the current quark mass it is given by the 
Gell-Mann Oakes Renner (GOR) relation,
\beq
    m_\pi^2 \, f_\pi^2 \= -m_0 \,\ave{\pb\psi} \;.
\label{GOR}
\eeq
Our scheme is consistent with this relation up to next-to-leading order
in $1/N_c$. Expanding both sides of Eq.~(\ref{GOR}) we get
\beq
m_{\pi}^{2(0)} f_{\pi}^{2(0)} \+ m_{\pi}^{2(0)}\delta f_{\pi}^2 
\+ \delta m_{\pi}^2 f_{\pi}^{2(0)}
= -m_0 \,  \Big(\ave{\pb\psi}^{(0)} + \delta\ave{\pb\psi}\Big) \;.
\label{gor1}
\eeq
Here $m_{\pi}^{2(0)}$ denotes the leading-order contribution to the 
squared pion mass, while $\delta m_{\pi}^2$ is the next-to-leading
order contribution. For $f_\pi^2$ and $\ave{\pb\psi}$ we introduce 
similar notations.
Since the GOR relation holds only in lowest order in $m_0$, Eq.~(\ref{gor1})
corresponds to a double expansion: $m_\pi^2$ has to be calculated in
linear order in $m_0$, $f_\pi^2$ and $\ave{\pb\psi}$ in the chiral limit.

The quark condensate is given by
\beq
    \ave{\pb\psi} \= -i \intp Tr {\tilde S}(p) \;, 
    \label{qbarq}
\eeq
with the $1/N_c$-corrected quark propagator
\beq
    {\tilde S}(p) \= S(p) \+ S(p)\,(\Sigma^{(a)} \+ \Sigma^{(b)}(p))\,S(p)
    \;.
\eeq
Here $S(p)$ is the Hartree propagator and $\Sigma^{(a)}$ and $\Sigma^{(b)}$
are the self energy corrections shown in Fig.~\ref{fig2}.
Since we are interested in a strict $1/N_c$ expansion, it should not be
iterated. The r.h.s. of Eq.~(\ref{qbarq}) can be easily evaluated
and one obtains
\beq
    \ave{\pb\psi}^{(0)} + \delta\ave{\pb\psi} \=
     - \frac{m-m_0}{2g_s} \;-\; \frac{D_\sigma(0)\,\Delta}{2g_s}\;,
    \label{qbqend}
\eeq 
with $\Delta$ as defined in Eq.~(\ref{tdself}).

The pion decay constant $f_{\pi}$ is calculated from the one-pion to vacuum
axial vector matrix element.
Basically this corresponds to evaluating the mesonic polarization diagrams,
Fig.~\ref{fig3}, coupled to an external axial current and to a pion.
This leads to expressions similar to Eqs.~(\ref{pol0}) and (\ref{deltapi}),
but with one external vertex equal to $\gamma^\mu \gamma_5 \frac{\tau^a}{2}$,
corresponding to the axial current, and the second external vertex
equal to $g_{\pi qq} i\gamma_5 \tau^b$, corresponding to the pion.
Here the $1/N_c$-corrected pion-quark coupling constant is 
defined as
\beq
    g_{\pi qq}^{-2} \= g_{\pi qq}^{-2(0)} \+ \dg \=
\frac{d{\tilde\Pi}_\pi(q)}{dq^2}|_{q^2 = m_\pi^2} \;,
\label{gpiqq}
\eeq
analogously to Eq.~(\ref{mesonmass0}). Now we take the divergence of the
axial current and then use the axial Ward identity 
\beq
\gamma_5 \,\pslash \= 2 m \gamma_5 \,+ \gamma_5 \,S^{-1}(k+p) 
                   \,+ S^{-1}(k)\, \gamma_5
\label{axialward}
\eeq
to simplify the expressions~\cite{dmitrasinovic}. One finds:
\beq
    f_{\pi} \= g_{\pi qq} \; \Big(\; \frac{{\tilde \Pi}_\pi(q) - 
    {\tilde \Pi}_\pi(0)}{q^2}\,m \+ \frac{\Pi_\pi(q) - \Pi_\pi(0)}{q^2} 
    \, D_\sigma(0)\,\Delta\; \Big)\Big|_{q^2=m_\pi^2}\;. 
\label{fpi}
\eeq
In the chiral limit, $q^2 = m_\pi^2\rightarrow 0$,
Eqs.~(\ref{mesonmass0}) and (\ref{gpiqq}) can be employed to replace 
the difference ratios on the r.h.s. by pion-quark coupling 
constants.
When we square this result and only keep the leading order and the
next-to-leading order in $1/N_c$ to finally obtain:
\beq
f_{\pi}^{2(0)} \+ \delta f_{\pi}^2 \= 
m^2 \,g_{\pi qq}^{-2(0)} \+ \Big(\,m^2 \,\dg  \+
2m\,D_\sigma(0)\,\Delta\,g_{\pi qq}^{-2(0)}\,\Big)~.
\label{fpi2}
\eeq

For the pion mass we start from Eqs.~(\ref{dmtilde}) and (\ref{mesonmass1}) 
and expand the inverse pion propagator around $q^2=0$:
\beq
    1 \;-\; 2g_s\,{\tilde\Pi}_\pi(0) \;-\; 2g_s\, 
    \Big(\frac{d}{dq^2}\,{\tilde\Pi}_\pi(q)\Big)\Big|_{q^2=0}\,m_\pi^2 
    \+ {\cal O}(m_\pi^4) \= 0 \;.  
\eeq
To find $m_\pi^2$ in lowest non-vanishing order in $m_0$ we have to
expand $1-2g_s {\tilde\Pi}_\pi(0)$ up to linear order in $m_0$, while
the derivative has to be calculated in the chiral limit, where it can
be identified with the inverse squared pion-quark coupling constant,
Eq.~(\ref{gpiqq}). The result can be written in the form
\beq
    m_\pi^2 \= \frac{m_0}{m}\,\frac{g^2_{\pi qq}}{2g_s}\;
    \Big(\,1 \,-\,\frac{D_\sigma(0)\Delta}{m}\,\Big) 
    \+  {\cal O}(m_0^2) \;.
\label{mpi}   
\eeq
Finally one has to expand this equation in powers of $1/N_c$.
This amounts to expanding $g_{\pi qq}^2$, which is the only term in 
Eq.~(\ref{mpi}) which is not of a definite order in $1/N_c$.
One gets:
\beq
    m_\pi^{2(0)} \+ \delta m_\pi^2 \= 
    m_0 \,\frac{m}{2g_s}\,\frac{g^{2(0)}_{\pi qq}}{m^2} \;-\;
    m_0 \,\frac{m}{2g_s}\,\frac{g^{2(0)}_{\pi qq}}{m^2} 
    \Big(\,g_{\pi qq}^{2(0)}\,\dg 
    \,+\,\frac{D_\sigma(0)\Delta}{m}\,\Big) 
    \;.
\label{mpinc}   
\eeq
It can be seen immediately that the leading-order term is exactly equal to 
$-m_0 \ave{\pb\psi}^{(0)}/f_\pi^{2(0)}$, as required by the GOR relation. 
Moreover, combining Eqs.~(\ref{qbqend}) for $m_0$~=~0, (\ref{fpi2}) and 
(\ref{mpinc}) one finds that the GOR relation in next-to-leading order, 
Eq.~(\ref{gor1}), holds in our scheme.

However, one should emphasize that this result is obtained by a strict 
$1/N_c$-expans\-ion of the various properties which enter into the GOR
relation and of the GOR relation itself. If one takes $f_\pi$ and $m_\pi$
as they result from Eqs.~(\ref{fpi}) and (\ref{mpi}) and inserts them 
into the l.h.s. of Eq.~(\ref{GOR}) one will in general find deviations from 
the r.h.s. which are due to higher-order terms in $1/N_c$.  
In this sense one can take the  violation of the GOR relation as a measure 
for the importance of these higher-order terms \cite{oertel}.

\section{Relation to hadronic models}
\label{hadron}
This section discusses an approximation to our scheme which 
points out the relation to hadronic models. 
In order to suppress the quark effects in the present model, it is
suggestive to assume that the constituent quark mass is very large as compared to the
relevant meson momenta. This assumption leads to an effective zero momentum 
approximation, i.e. the quark vertices are taken at zero incoming momentum
(``static limit''). 
In order to preserve chiral symmetry we then have to approximate the RPA-meson 
propagators consistently. This amounts to replacing the function $I(p)$ 
in the RPA polarization functions by $I(0)$. 
The latter can be related to the leading-order pion decay constant 
in this approximation.
The RPA-meson propagators then take the form of free boson propagators:
\beq
D^{(0)}_M(q) = \frac{g^{2(0)}_{Mqq}}{q^2-m^{(0)2}_M}~,
\eeq
with
\beq
    m_\pi^{2(0)} = \frac{m_0 m}{2g_s f_\pi^{2(0)}}~, \qquad
    m_\rho^{2(0)} = \frac{3 m^2}{4g_v f_\pi^{2(0)}}~, \qquad
    m_\sigma^{2(0)} = m_\pi^{2(0)} + 4m^2~, \quad
    m_{a_1}^{2(0)} = m_\rho^{2(0)} + 6m^2~,
\label{mstat}
\eeq
and
\beq
g^{2(0)}_{\pi qq} \= g^{2(0)}_{\sigma qq} \= \frac{2}{3} g^{2(0)}_{\rho qq}
\= \frac{2}{3} g^{2(0)}_{a_1 qq} \= \frac{m^2}{f_\pi^{2(0)}} \;.
\eeq
Next we expand the quark triangles and box diagrams to first order
in the external momenta.
In this way one obtains a hadronic model with effective meson-meson coupling
constants.  Let us, for example, look at the $\rho$-meson self-energy
which is generated as an approximation to the $1/N_c$-corrected NJL model. 
As the $\rho\rho\sigma$- and the $\rho\rho\sigma\sigma$-vertex vanish in this
approximation, we are left with a pion loop diagram, which is generated from
the diagram shown in Fig.~\ref{fig2}(a), and a pion tadpole diagram, which
arises from the sum of the diagrams in Fig.~\ref{fig2}(b) and (c). These are
exactly the diagrams which are calculated in standard hadronic descriptions 
for the $\rho$-meson in vacuum, e.g. \cite{herrmann,chanfray}.

In hadronic models one is also interested in the coupling of a bare
$\rho$-meson to a photon, which is e.g. needed to calculate the pion
electromagnetic form factor via vector-meson dominance. In the NJL model the 
bare $\rho$-meson corresponds to the RPA meson and its coupling to a photon
is basically given by the RPA polarization loop in the vector channel.
If we now perform the same low-momentum approximations as for the 
RPA-meson propagators we find that the vertex is given by
$-i (e/g_{\rho\pi\pi}) q^2 g_{\mu\nu}$
which exactly corresponds to the $\gamma\rho$ vertex in the vector
dominance model of Kroll, Lee and Zumino \cite{klz}. 

For the effective $\rho\pi\pi$ coupling constant we obtain 
\beq
g^2_{\rho\pi\pi} = 6 g^{2(0)}_{\pi qq} = 6 \frac{m^2}{f^{(0)2}_{\pi}}~.
\label{rhopipistat}
\eeq
This result can be used to estimate the importance of quark effects.  
If we take the commonly used value of about 6 for $g_{\rho\pi\pi}$ in 
hadronic models we obtain for the constituent quark mass
\beq
    m \approx \sqrt{6} f^{(0)}_{\pi}~.
\eeq 
As will be seen in the next section the leading-order pion decay constant
is larger but not much larger than the $1/N_c$-corrected quantity. 
If we take  $f_{\pi}^{(0)} \sim$ 100-150 MeV, we find that the constituent 
quark mass should be about 250-350 MeV, which is obviously in contradiction 
to our original assumption of very heavy quarks . Thus we would expect
that this approximation does not describe the full model very well. This 
point will be discussed in more detail in the next section. 

Similar to the $\rho$-meson one can perform approximations to the
$1/N_c$-corrected self-energies of the other mesons.
The authors of Ref.~\cite{davesne} have used effective meson-meson
coupling constants generated from this approximation to the NJL quark loops
in a linear sigma model calculation for in-medium pion properties. It
is nice to see that a consistent approximation to the $1/N_c$-corrected NJL
model naturally generates an effective one loop approximation to the linear
sigma model in the $\pi-\sigma$ sector. 

A slightly different approximation to the NJL model has been performed
in ref.~\cite{he}, where instead of a low-momentum expansion, the
vertices are evaluated for on-shell intermediate mesons. 
However, at least for processes dominated by intermediate pions
this gives very similar results to those obtained with the low-momentum
expansion.

\section{Numerical results}
\label{numerics}
In this section we present numerical results for the self-energy
of the $\rho$-meson and related quantities, such as the electromagnetic
form factor of the pion and $\pi$-$\pi$ phase shifts in the 
vector channel.  Before we begin with the explicit calculation we have to
come back to the regularization. As discussed in
Sec.~\ref{pion}, all quark loops, i.e. the RPA polarization diagrams,
the quark triangles and the quark box diagrams should be regularized in the 
same way in order to preserve chiral symmetry.
We use a Pauli-Villars-regularization with two regulator masses 
${\sqrt{m^2+\Lambda_q^2}}$ and ${\sqrt{ m^2 +2 \Lambda_q^2}}$.
As before, $m$ denotes the constituent quark mass and $\Lambda_q$ is the 
cutoff parameter. The regularization of the meson loop
(integration over $d^4p$ in eq.~(\ref{deltapi})) is not constrained by 
chiral symmetry and independent from the quark loop regularization. 
For practical reasons we choose a three-dimensional cutoff $\Lambda_M$ 
in momentum space. In order to get a well-defined result we work in the 
rest frame of the $1/N_c$-improved meson.
The same regularization scheme was already used
in Ref.~\cite{oertel}. 

With the additional meson cutoff $\Lambda_M$ there are five
parameters: the current quark mass $m_0$, the two coupling constants
$g_s $ and $g_v$, the quark-loop cutoff $\Lambda_q$ and the meson-loop
cutoff $\Lambda_M$. 
For a given value of $\Lambda_M$, the current quark mass $m_0$, the scalar 
coupling constant $g_s$ and the cutoff $\Lambda_q$ can be fixed
by fitting the pion mass $m_{\pi}$, the pion decay constant $f_{\pi}$ 
and the quark condensate $\qq$ to their empirical values.
Then we can try to determine the two remaining parameters, i.e. 
the vector coupling constant $g_v$ and the meson cutoff $\Lambda_M$,  
by fitting the pion electromagnetic form factor in the time-like region,
which is related, via vector meson dominance, to the $\rho$-meson propagator 
(see below). Roughly speaking this amounts to fitting the $\rho$-meson mass 
and its width.
The $\rho$-meson mass is most sensitive to the coupling constant $g_v$. 
Since in the present article we neglect possible $\rho$ and $a_1$ meson 
intermediate states, the value of $g_v$ does not influence the observables 
in the pion sector, $m_{\pi}, f{\pi}$ and $\qq$.
Hence for given $\Lambda_M$ we can try to fit these observables together
with the $\rho$-meson mass. Finally, we determine $\Lambda_M$ by 
comparing the resulting pion electromagnetic form factor with the
experimental data. 

However, this procedure contains a problem: 
For $m_0 \neq 0$ the corrected pion mass $m_\pi$ is always larger than the 
RPA pion mass $m_\pi^{(0)}$. Hence, if we perform a fit for 
the $1/N_c$-improved pion mass, the masses of the RPA pions, which enter into the
intermediate states, e.g. in diagram Fig.~\ref{fig3}(a), are too 
small, i.e. the value of the threshold energy for the decay of the 
$\rho$-meson into two pions is too small.
Therefore, since we are mainly interested in the $\rho$-meson in this 
article, we adjust $m_{\pi}^{(0)}$, rather than $m_\pi$ to the physical
pion mass.    
Here we make use of the fact that the pion mass is very sensitive to the
current quark mass $m_0$ whereas the two other observables we want to
fit, $f_\pi$ and $\qq$, depend only weakly on $m_0$.  

Five parameter sets (corresponding to five different meson cutoff values 
$\Lambda_M$) are listed in Table~\ref{table1}, together with the
constituent quark mass $m$, the values of $m_\pi$, $f_\pi$ and $\qq$
and the corresponding leading-order quantities. As outlined above,
$\Lambda_q$, $m_0$ and $g_s$ have been obtained by fitting $m_{\pi}^{(0)}$, 
$f_\pi$ and $\qq$ to the empirical values.  
For the quark condensate this is not known very precisely, but the 
absolute value is probably below 2(260~MeV)$^3$.
(This corresponds roughly to the upper limit extracted in ref.~\cite{dosch} 
from sum rules at a renormalization scale of 1~GeV. Recent lattice 
results are $\qq$~=~-2($(231 \pm 4 \pm 8 \pm 6)$~MeV)$^3$ \cite{giusti}.)
It turns out that we can only stay below this limit, if the meson cutoff
is not too large ($\Lambda_M \lesssim$~700~MeV). 
This is related to the fact that the constituent quark mass and hence the
absolute value of the leading-order condensate $\qq^{(0)}$ increases
with $\Lambda_M$ if we keep $f_\pi$ constant.

Table~\ref{table1} also displays the ratio $-m_0 \qq / m_\pi^2 f_\pi^2$,
which would be equal to 1 if the GOR relation was exactly fulfilled.
Note that for all parameter sets given in the table the deviations are
less than 10\% (for $\Lambda_M \leq$~600~MeV 
even less than 3\%), indicating that higher-order corrections in $1/N_c$
are small.
\begin{table}[t]
\begin{center}
\begin{tabular}{|c|c|c|c|c|c|}
\hline
$\Lambda_M$~/~MeV   &   0.  & 300.  & 500.  & 600.  & 700.  \\ \hline
$\Lambda_q$~/~MeV   & 800.  & 800.  & 800.  & 820.  & 852.  \\ \hline
$m_0$~/~MeV         & 6.13  & 6.40  & 6.77  & 6.70  & 6.54  \\ \hline
$g_s\Lambda_q^2$    & 2.90  & 3.07  & 3.49  & 3.70  & 4.16  \\ \hline
$g_v$~/~$g_s$       &  --   &  --   & 1.0   & 1.6   & 2.4   \\ \hline
$m$~/~MeV           & 260.  & 304.  & 396.  & 446.  & 550.  \\ \hline
$m_\pi^{(0)}$~/~MeV & 140.0 & 140.0 & 140.0 & 140.0 & 140.0 \\ \hline
$m_\pi$~/~MeV       & 140.0 & 143.8 & 149.6 & 153.2 & 158.1 \\ \hline
$f_\pi^{(0)}$~/~MeV &  93.6 & 100.6 & 111.1 & 117.0 & 126.0 \\ \hline
$f_\pi$~/~MeV       &  93.6 &  93.1 &  93.0 &  93.1 & 93.4  \\ \hline
$\qq^{(0)}$~/~MeV$^3$ & -2(241.1)$^3$ & -2(249.3)$^3$ & -2(261.2)$^3$ 
                      & -2(271.3)$^3$ & -2(287.2)$^3$ \\ \hline
$\qq$~/~MeV$^3$       & -2(241.1)$^3$ & -2(241.7)$^3$ & -2(244.1)$^3$ 
                      & -2(249.5)$^3$ & -2(261.4)$^3$ \\ \hline
$\ave{r_\pi^2}^{1/2}$~/~fm & -- & -- & 0.59
                           & 0.61 & 0.66 \\ \hline
-$m_0 \qq / m_\pi^2 f_\pi^2$ & 1.001 & 1.007 & 1.018 & 1.023 & 1.072
\\ \hline
\end{tabular}
\end{center}
\caption{{\it The model parameters ($\Lambda_M$, $\Lambda_q$, $m_0$, 
$g_s$ and $g_v$) and the resulting values of $m_\pi$, $f_\pi$ and
$\qq$ (together with the corresponding leading-order quantities), 
the constituent quark mass $m$ and the charge radius of the pion
$\ave{r_\pi^2}^{1/2}$. The ratio $-m_0 \qq / m_\pi^2 f_\pi^2$,
is also given.
}}
\label{table1}
\end{table}

We now turn to the $\rho$-meson channel. 
According to Eq.~(\ref{pol1}), the polarization function of the $\rho$-meson 
is a sum of the RPA polarization loop and the four $1/N_c$-correction terms, 
as shown in Fig.~\ref{fig3}:
\beq
    {\tilde \Pi}_\rho^{\mu\nu, ab}(q) \= \Pi_\rho^{\mu\nu, ab}(q)
\+ \sum_{k=a,b,c,d}\; \delta \Pi_\rho^{(k)\;\mu\nu, ab}(q)
\;.
\label{rhopol1}
\eeq
Because of vector current conservation, the polarization function 
has to be transverse, i.e.
\beq
    q_\mu\,{\tilde \Pi}_\rho^{\mu\nu, ab}(q) \=
    q_\nu\,{\tilde \Pi}_\rho^{\mu\nu, ab}(q) \= 0 \;.
\label{trans}
\eeq
For our scheme it can be shown with the help of Ward identities
that these relations hold, if we assume that the regularization  
preserves this property.
This is the case for the Pauli-Villars regularization scheme, which was 
employed to regularize the RPA part ${\Pi}_\rho$. 
Together with Lorentz covariance this leads to Eq.~(\ref{pirho}) for
the tensor structure of ${\Pi}_\rho$.
On the other hand, since we use a three-dimensional sharp cutoff for the 
regularization of the meson loops, the $1/N_c$-correction terms are in
general not transverse. However, as mentioned above, we work in the
rest frame of the $\rho$-meson, i.e. ${\vec q}$~=~0.
In this particular case Eq.~(\ref{trans}) is not affected by the cutoff 
and the entire function ${\tilde \Pi}_\rho$ can be written in the form of 
Eq.~(\ref{pirho}):

\beq
    {\tilde \Pi}_\rho^{\mu\nu, ab}(q) \= 
    {\tilde \Pi}_\rho(q)\,T^{\mu\nu}\, \delta_{ab} \=
    \Big(\Pi_\rho(q) \+ \sum_{k=a,b,c,d}\; \delta \Pi_\rho^{(k)}(q) \Big)
    \,T^{\mu\nu}\, \delta_{ab}
    \;,
\label{pirho1}
\eeq
i.e. instead of evaluating all tensor components separately
we only need to calculate the scalar functions
$\Pi_\rho = -1/3\,g_{\mu\nu}\,\Pi_\rho^{\mu\nu}$ and 
$\delta\Pi_\rho^{(k)} = -1/3\,g_{\mu\nu}\,\delta\Pi_\rho^{(k)\,\mu\nu}$.  

A second consequence of vector current conservation is, that the 
polarization function should vanish for $q^2=0$.
For the $1/N_c$-correction terms this is violated by the sharp cutoff.
We cure this problem by performing a subtraction:
\beq
    \sum_{k=a,b,c,d}\; \delta \Pi_\rho^{(k)}(q) \;\rrr\;
    \sum_{k=a,b,c,d}\; \Big(\delta \Pi_\rho^{(k)}(q) \;-\;
    \delta \Pi_\rho^{(k)}(0) \Big) \;.
\label{sub}
\eeq 
Note, however, that already at the RPA level a subtraction is required,
although the RPA part is regularized by Pauli-Villars.
This is due to a rather general problem which is discussed in detail in
App.~\ref{correlators}.

\begin{figure}[t]
\hspace{1cm}
\parbox{14cm}{
     \epsfig{file=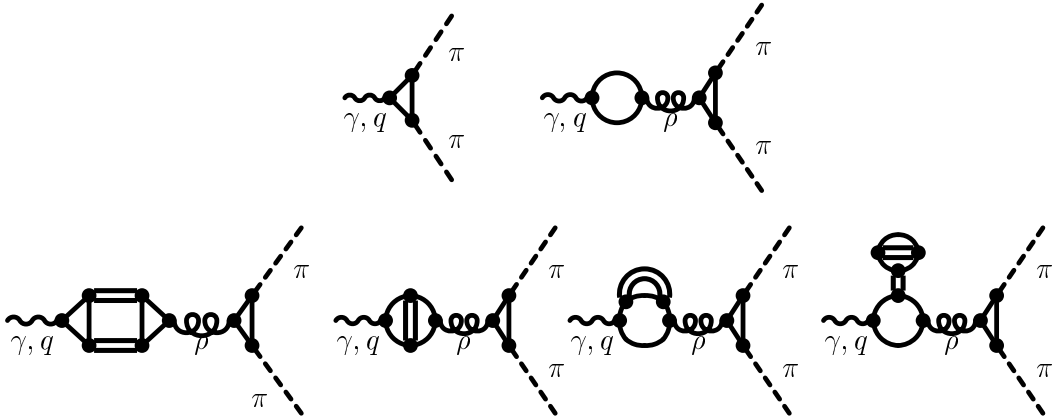,
     height=5.6cm, width=14.cm}}
\caption{\it Contributions to the pion electromagnetic form factor.
         Note that the propagator denoted by the curly line corresponds
         to the $1/N_c$-corrected $rho$-meson, while the double lines
         indicate RPA pions and sigmas.}
\label{fig5} 
\end{figure}

As mentioned above, we want to fix the remaining two model parameters,
i.e. $g_v$ and $\Lambda_M$, by fitting the pion electromagnetic form factor,
$F_\pi(q)$, in the time-like region, which is dominated by the 
$\rho$-meson. The diagrams we include in our calculations are shown in
Fig.~\ref{fig5}. 
The two diagrams in the upper part correspond to the standard
NJL description of the form factor \cite{lutz} if the full
$\rho$-meson propagator (curly line) is replaced by the RPA one.
Hence, the first improvement is the use of the $1/N_c$-corrected 
$\rho$-meson propagator in our model. Since, in the standard scheme,
the photon couples to the $\rho$-meson via a quark-antiquark polarization 
loop, in our scheme we should also take into account the $1/N_c$-corrections
to the polarization diagram for consistency. This leads to the diagrams
in the lower part of Fig.~\ref{fig5}. On the other hand the external
pions are taken to be RPA pions (i.e. mass $m_\pi^{(0)}$ and 
pion-quark-quark coupling constant $g_{\pi qq}^{(0)}$). 
This is more consistent with the fact that the $\rho$-meson is also dressed 
by RPA pions and, as discussed above, we have fitted $m_\pi^{(0)}$ 
to the experimental value.

Evaluating the diagrams of Fig.~\ref{fig5} we obtain for the form factor  
\beq
|F_{\pi}(q)|^2 = \frac{1}{2}\,\Big|\,g_{\pi qq}^{(0)2} 
f(p,p^{\prime}) \, \Big(1-{\tilde\Pi}_{\rho}(q) 
{\tilde D}_{\rho}(q)\Big)\,\Big|^2~,
\label{piform}
\eeq  
where $p$ and $p'$ are the four-momenta of the external pions and
$f(p,p^{\prime})$ is a scalar function appearing in the
$\pi\pi\rho$-vertex function (see App.~\ref{functions}).
For the form factor it has to be evaluated for on-shell pions, i.e. 
$p^2 = p^{\prime2} = m_{\pi}^2$ and $p\cdot p^{\prime} = q^2/2-m_{\pi}^2$.  
From its explicit form, which is given in App.~\ref{functions}, one can 
show that $g_{\pi qq}^{(0)2} f(p,p^{\prime}) = \sqrt{2}$ for $q^2 = 0$.
Since the $\rho$-meson self energy ${\tilde\Pi}_{\rho}$ vanishes at this 
point, we find $F_\pi(0) = 1$, as it should be.  
\begin{figure}[t]
\parbox{6cm}{
     \epsfig{file=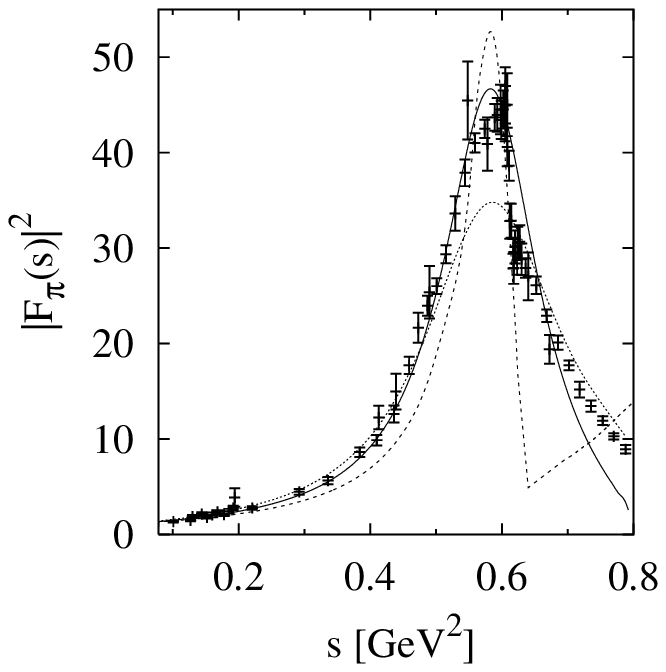,
     height=6cm, width=7.cm}\quad}
\hspace{2cm}
\parbox{6cm}{
     \epsfig{file=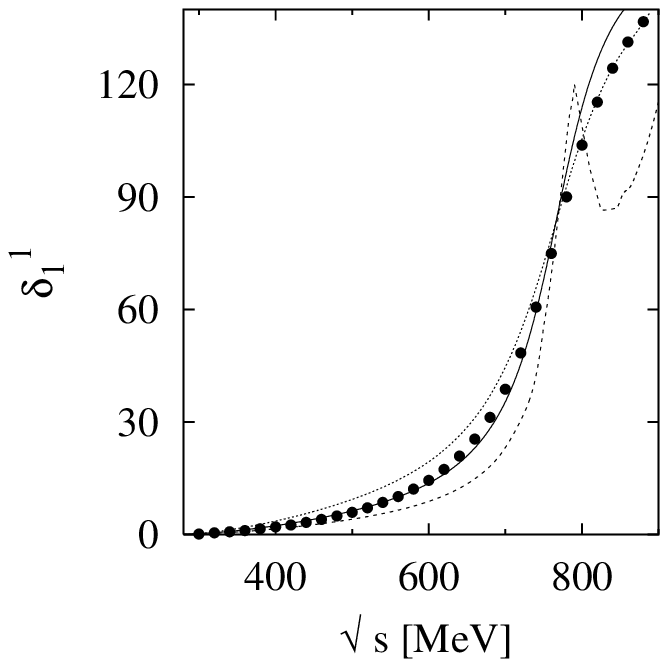,
     height=6cm, width=7cm}\quad}
\hspace{1cm}
\caption{\it The pion electromagnetic form factor (left panel) and 
             the $\pi\pi$-phase 
             shifts in the vector-isovector channel (right panel) for different meson 
             cutoff parameters, $\Lambda_M=500$ MeV (dashed), $\Lambda_M=600$ MeV 
             (solid) and $\Lambda_M=700$ MeV (dotted). 
             The data points are taken from refs.~\cite{barkov} and
             \cite{frogatt}, respectively.}
\label{fig6} 
\end{figure}

The numerical results for $|F_\pi|^2$ as a function of the center-of-mass
energy squared are displayed in the left panel of Fig.~\ref{fig6}, 
together with the experimental data \cite{barkov}. 
The various curves correspond to
different values of the meson cutoff $\Lambda_M$.  
For the other parameters the values listed in Table~\ref{table1}
are taken. The vector coupling constant $g_v$ is chosen such that the 
maximum of the form factor is at the correct position. 
Roughly speaking this corresponds to fitting the mass of the (dressed) 
$\rho$-meson.    
We only show the results for $\Lambda_M \geq$~500~MeV. For lower 
meson cutoffs the unphysical quark-antiquark decay threshold of the 
$\rho$-meson is below the maximum of the form factor and makes a
comparison not very meaningful.
 
For $\Lambda_M$~=~500~MeV (dashed line) the quark-antiquark threshold
is at $s$~=~0.63~GeV$^2$, causing a cusp in the form factor
slightly above the maximum. Because of the sub-threshold attraction in
the $\rho$-meson channel the form factor drops very steeply below the
cusp, leading to a poor description of the data above the maximum. 
In addition, also at the rising edge of the peak, where
threshold effects are less important, we see that the width of the
form factor is underestimated by the calculation, i.e. $\Lambda_M$
is too small. On the other hand, a meson cutoff of 700~MeV (dotted line) 
is already too large. In particular, the height of the maximum is
underestimated.   
With $\Lambda_M$~=~600~MeV (solid line) we obtain the best description
of the data. 
Since we assumed exact isospin symmetry in our model we can, of course,  
not reproduce the detailed structure of the form factor around 
0.61~GeV$^2$, which is due to $\rho$-$\omega$-mixing.  
The high-energy part above the peak is somewhat underestimated,
mainly due to the $q\bar q$-threshold at $s$~=~0.80~GeV$^2$.
Probably the fit can be somewhat improved if we take a slightly 
larger meson cutoff, but we are not interested in fine-tuning here.
In addition, it is to be expected that the inclusion of $\rho$-
and $a_1$ intermediate states will improve the high-energy behavior. 

A closely related quantity is the charge radius of the pion,
which is defined as
\beq
     \ave{r_\pi^2} \= 6\,\frac{d F_\pi}{d q^2}\Big|_{q^2 = 0} \;.
\eeq 
Results for $\Lambda_M$~=~500, 600 and 700~MeV are listed in 
Table~\ref{table1}. All of them are close to the experimental value, 
$\ave{r_\pi^2}^{1/2}$~=~(0.663~$\pm$~0.006)~fm \cite{amendolia}.
For $\Lambda_M$~=~700~MeV we find perfect agreement. 
On the other hand, a simple pole ansatz for the form factor leads to
$\ave{r_\pi^2}^{1/2} = \sqrt{6}/m_\rho$~=~0.63~fm \cite{bhaduri}, 
which is of the same quality as our results. Obviously, at $q^2$~=~0 
we are not very sensitive to the details of the $\rho$-meson peak,
while other effects which are not included in our model might start
to play a role.
Therefore the pion charge radius is certainly not very well suited
for fixing our model parameters. 

\begin{figure}[h]
\hspace{3.5cm}
     \epsfig{file=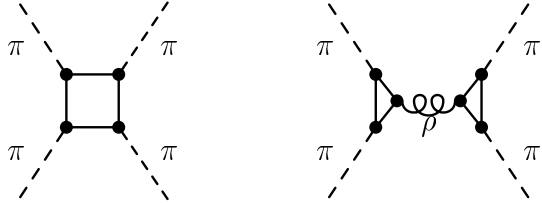}
\caption{\it Diagrams contributing to the $\pi\pi$-scattering 
             amplitude: Quark box diagram (left) and 
             s-channel $\rho$-meson exchange (right).}
\label{fig7} 
\end{figure}

One can also look at the $\pi\pi$-phase shifts in the vector-isovector
channel. We include the diagrams shown in Fig.~\ref{fig7}, i.e.
the s-channel $\rho$-meson exchange and the direct $\pi\pi$-scattering
via a quark box diagram. The latter has to be projected onto spin and
isospin~1, which is a standard procedure. (For example, the 
analogous projection onto spin and isospin~0 can be found in 
Refs.~\cite{davesne,bernard}.)  
Our results, together with the empirical data \cite{frogatt}, are
displayed in the right panel of Fig.~\ref{fig6}. 
Since the main contribution comes from the s-channel $\rho$-meson 
exchange, they more or less confirm our findings for the form factor:
below the $\rho$-meson peak the best fit of the data is obtained with
$\Lambda_M$~=~600~MeV while, at higher energies, where $q\bar q$-threshold 
effects start to play a role, the data are better described with 
$\Lambda_M$~=~700~MeV. 

Finally, we wish to compare the results of the full $1/N_c$-corrected
NJL model with the static limit, i.e. the approximation introduced in 
Sec.~\ref{hadron}.
As pointed out, this approximation corresponds to a purely hadronic
description where all quark effects are suppressed.  
For instance, taking the static limit of the diagrams shown 
in Fig.~\ref{fig5}, we exactly recover the diagrams which contribute to the 
pion electromagnetic form factor in the hadronic model of ref.~\cite{klingl}. 
On the other hand, we already estimated that quark effects should not
be negligible, i.e. we expect the static limit not to be a good 
approximation to the exact NJL calculations.  
In fact, if we take the parameters of our best fit to the pion 
electromagnetic form factor ($\Lambda_M$~=~600~MeV) and insert the
corresponding values for $m$ and $f_\pi^{(0)}$ (see Table~\ref{table1})
into Eq.~(\ref{rhopipistat}) we find $g_{\rho\pi\pi}$~=~9.3, which is
considerably larger than the value of $\sim 6$,  typically needed
in hadronic models to describe the data. 
\begin{figure}[t]
\hspace{4cm}
\parbox{6cm}{
     \epsfig{file=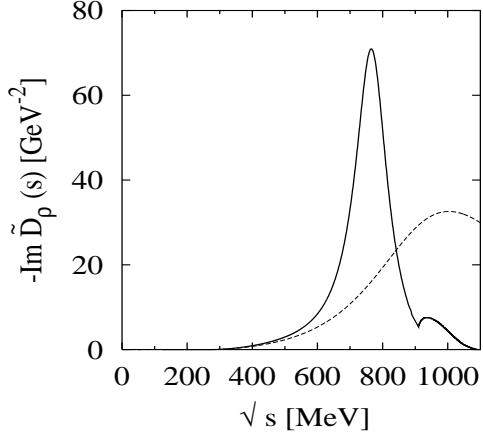,
     height=6cm, width=7.cm}\quad}
\hspace{4cm}
\caption{\it The imaginary part of the $1/N_c$-corrected $\rho$-meson
             propagator. The solid line represents the full calculation 
             while the dashed line denotes the static limit.}
\label{fig8} 
\end{figure}
Fig.~\ref{fig8} displays the imaginary part of the $\rho$-meson 
propagator for $\Lambda_M$~=~600~MeV. 
The solid line indicates the result of an exact treatment of our 
model. Since the parameters have been fitted to the pion electromagnetic 
form factor, the maximum is close to the empirical $\rho$-meson 
mass of 770~MeV. The cusp at $\sqrt{s}$~=~892~MeV is again a 
$q\bar q$-threshold effect. 
The corresponding result in the static limit is displayed by 
the dashed line. Of course there is no $q\bar q$-threshold in this 
approximation. Obviously the peak is much broader and shifted to higher 
energies.

\begin{figure}[t!]
\parbox{6cm}{
     \epsfig{file=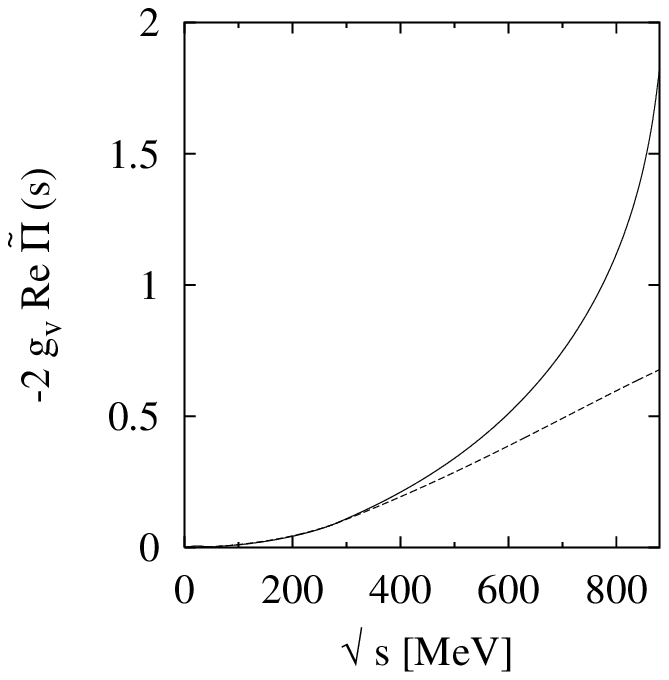,
     height=6cm, width=7.cm}\quad}
\hspace{2cm}
\parbox{6cm}{
     \epsfig{file=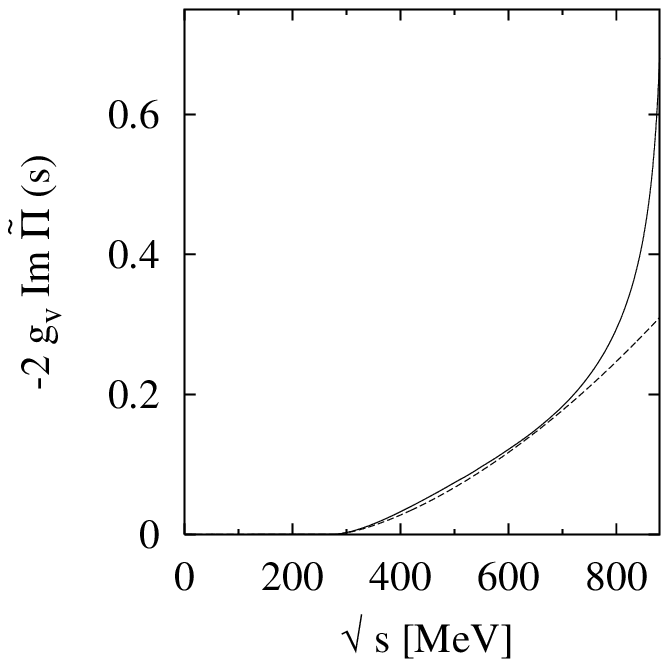,
     height=6cm, width=7cm}\quad}
\hspace{1cm}
\caption{\it The real part (left) and the imaginary part (right) of the
             $1/N_c$-corrected $\rho$-meson self-energy, multiplied
             by $-2g_v$. The solid line represents the full 
             calculation, while the dashed line indicates the static limit.}
\label{fig9} 
\end{figure}

For a better understanding of this behavior we also compare the real-
and imaginary parts of the corresponding $\rho$-meson self-energies 
(Fig.~\ref{fig9}). Obviously, the differences seen in Fig.~\ref{fig8}
are mainly due to the real part which is much more attractive in the 
exact NJL calculation than in the static limit, while the 
imaginary parts are not very different up to $\sqrt{s} \sim$~800~MeV.
Note that, in this region, the imaginary part exclusively results from
the $\rho\rightarrow\pi\pi$ process, i.e. quarks do not show up as 
unphysical decay channels. The differences to the static limit,
both in the real- and in the imaginary part, come 
about by the fact, that the quarks cause a non-trivial momentum 
dependence of the intermediate meson propagators and of the effective 
meson-meson vertices. For instance, the quark triangles do not exactly 
behave as point-like three-meson vertices. 
This might be viewed as ``physical quark effects'' in contrast to the
unphysical $q\bar q$-decays at higher energies.
However, the momentum dependence of any quark loop below the 
$q\bar q$-threshold is related to the imaginary part above the threshold 
via dispersion relations. 
In that sense any difference to the static limit could be interpreted as
an unphysical effect and one might ask whether ``physical quark effects''
exist at all.  
Of course, this fundamental question is far beyond the scope of our paper
and can probably not be answered without understanding the mechanism of 
quark confinement itself. 
On the other hand, at sufficiently low energies ($\lsim$~400~MeV) the exact
self energy of the $\rho$ meson does indeed almost coincide with the static 
limit. Here one can nicely see, how a hadronic model, which incorporates
chiral symmetry, emerges naturally from the underlying quark structure once
the many-body theory is carried to a sufficient degree of sophistication.  

\section{Conclusions}
\label{conclusions}
We have investigated meson properties within the Nambu--Jona-Lasinio model, 
including meson-loop corrections, which are generated via a systematic 
$1/N_c$-expansion of the self energy in next-to-leading order. 
We have shown that such a scheme is consistent with chiral symmetry, 
leading to massless pions in the chiral limit. For non-vanishing current 
quark masses the pion mass is consistent with the Gell-Mann Oakes Renner 
relation if one carefully expands both sides of the relation up to 
next-to-leading order in $1/N_c$.

The relative importance of the $1/N_c$-corrections is controlled by a 
parameter $\Lambda_M$, which cuts off the three-momenta of the meson loops. 
One of the main goals of the present article was to determine the value of 
$\Lambda_M$, together with the other parameters of the model. 
To that end we have performed a fit of the quark condensate $\qq$ and the
pion decay constant $f_\pi$, together with the pion electromagnetic
form factor $F_\pi$ in the time-like region. The latter more or less 
amounts to fitting the mass and the width of the $\rho$-meson. 
Here the meson loops are absolutely crucial in order to include the 
dominant $\rho \rightarrow \pi\pi$-decay channel, while the leading-order 
approximation contains only unphysical $q\bar q$-decay channels.
Of course, a priori it was not clear to what extent these unphysical
decay modes, which are an unavoidable consequence of the missing
confinement mechanism in the NJL model, are still present in the region 
of the $\rho$-meson peak. 

It turns out that a reasonable fit of the above observables can be 
achieved with $\Lambda_M$~=~600~MeV. 
For the constituent quark mass we find $m$~=~446~MeV. 
Hence the unphysical $q\bar q$-decay channel opens at 892~MeV, 
about 120~MeV above the maximum of the $\rho$-meson peak. 
Our result is also interesting in the context of Ref.~\cite{oertel}
where we have reported on the existence of instabilities in the pion sector
for very large values of $\Lambda_M$. However, with $\Lambda_M$~=~600~MeV,
as determined by our parameter fit, we are far away from this region.
In fact, we find only moderate changes in the pion and quark sector:
$f_\pi$ and $\qq$ are lowered by about 20\% by the meson loop
corrections, while the pion mass is increased by about 10\%. This
also indicates that our scheme converges rapidly and higher-order
terms in the $1/N_c$-expansion are small. 

We have discussed that a hadronic model can be derived from our model 
if one takes the so-called static limit. Basically, this corresponds 
to a low-momentum approximation to the quark loops. 
In this way one can study the importance of quark effects, which
are present in the exact model because of lack of confinement, 
but absent in the static limit.
At higher energies we find for the $\rho$-meson rather large differences
between the exact propagator and the static limit, whereas at sufficiently low
energies ($\lsim$~400~MeV) the exact calculation and the static limit almost
coincide. This implies that in the low-energy region the unphysical quark effects are
suppressed and demonstrates  that a hadronic model with realistic 
parameters indeed emerges from the underlying quark structure. 
It is likely that the quark threshold and hence the unphysical effects 
from continuum $\bar qq$ decay can be
pushed to even higher energies once intermediate $\rho$-and $a_1$
meson states are included. Given the successful vacuum description of the
$\rho$ meson, as presented in the present paper, it will be interesting
to extend the calculation to finite temperature to asses medium
modifications in the presence of a thermal heat bath. Because of
spontaneous chiral symmetry breaking and its restoration at high temperature
this implies a simultaneous treatment of the $a_1$ meson. Work in this
direction is in progress.

\section*{Acknowledgements}
We are indebted to G.J. van Oldenborgh for his assistance in questions
related to his program package FF (see  
{\it http://www.xs4all.nl/$\sim$gjvo/FF.html}),
which was used in parts of our numerical calculations.
We also thank M. Urban for illuminating discussions. 
This work was supported in part by the BMBF and NSF grant NSF-PHY98-00978.


\begin{appendix}

\section{Definition of elementary integrals}
\label{integrals}
It is possible to reduce the expressions for the quark loops to some
elementary integrals~\cite{passarinoveltman}, see App.~\ref{correlators}
and~\ref{functions}. In this section we give the definitions of these 
integrals. 
\bea
&&\hspace{-12mm} I_1 = \intk \frac{1}{k^2-m^2+i \eps}~, \label{onepoint}\\
&&\hspace{-12mm} I(p) = \intk \frac{1}{(k^2-m^2+i \eps)( (k+p)^2-m^2+
  i\eps)}~,\label{twopoint}\\
&&\hspace{-12mm} K(p) = \intk \frac{1}{(k^2-m^2+i \eps)^2( (k+p)^2-m^2+
  i\eps)}~,\label{threepoint1} \\
&&\hspace{-12mm} M(p_1,p_2) = \intk \frac{1}{(k^2-m^2+i \eps)
                 ( k_1^2-m^2+ i\eps)( k_2^2-m^2+i \eps)}~,\label{threepoint}\\
&&\hspace{-12mm} L(p_1,p_2,p_3) = \intk\frac{1}{(k^2-m^2+i \eps)
  ( k_1^2-m^2+ i\eps)
  ( k_2^2-m^2+i \eps)( k_3^2-m^2 + i\eps)}~,\label{fourpoint}\\
&&\hspace{-12mm} p_1^{\mu} M_1(p_1,p_2)+p_2^{\mu} M_1(p_2,p_1)
  = \intk \frac{k^{\mu}}{(k^2-m^2+i \eps)( k_1^2-m^2+ i\eps)
  (k_2^2-m^2+i\eps)},\label{tensor}
\eea
with $ k_i = k+p_i$. 
The function $M_1(p_1,p_2)$ can be expressed in terms of the other integrals:
\beq
M_1(p_1,p_2)
= \frac{p_1\!\cdot\!p_2\ I(p_1) - p_2^2\ I(p_2)
   + (p_2^2-p_1\!\cdot\!p_2)\  I(p_1-p_2)  
  + p_2^2\ (p_1^2-p_1\!\cdot\!p_2)\ M(p_1,p_2)}
  {2 \; ((p_1\!\cdot\!p_2)^2 \,-\, p_1^2\,p_2^2)}~,
\eeq
All integrals in Eqs.~(\ref{onepoint}) to (\ref{tensor}), are understood 
to be regularized. In our model we use Pauli-Villars regularization with 
two regulators, i.e. we replace
\beq
    \intk f(k;m) \;\rrr\; \intk \sum_{j=0}^2 c_j\,f(k;\mu_j)~,
    \label{pv}
\eeq
with
\beq
    \mu_j^2 \= m^2 \+ j\,\Lambda_q^2~;  \qquad
    c_0 = 1, \quad c_1 = -2, \quad c_2 = 1~.
\eeq
One then gets the following relatively simple analytic expressions for 
the integrals $I_1$, $I(p)$ and $K(p)$: 
\bea
&&\hspace{-12mm} 
  I_1=\frac{-i}{16\pi^2}\sum_j c_j\, \mu_j^2 \ln(\mu_j^2)\label{i1}\\
&&\hspace{-12mm}
  I(p) = \frac{-i}{16\pi^2}\sum_j c_j\, \Big(x_{j1} \ln(x_{j1})
  +x_{j2} \ln(-x_{j2})+x_{j1}\ln(-p^2x_{j1})+x_{j2}\ln(p^2x_{j2})\Big)\\ 
&&\hspace{-12mm}
  I(p=0) = \frac{-i}{16\pi^2}\sum_j c_j\, \ln(\mu_j^2)\label{i0}\\
&&\hspace{-12mm}
  K(p) = \frac{-i}{16\pi^2}\sum_j c_j\,\frac{1}{2 p^2(x_{j1}-x_{j2})} 
\Big(-\ln(x_{j1})-\ln(-x_{j1})+ \ln(x_{j2})+\ln(-x_{j2})\Big)~,
\eea
with 
\beq
x_{j1,2} = {1\over2}\pm{1\over2}\sqrt{1-{4 \mu_j^2\over p^2}} \;.
\eeq
An analytic expression for the three-point function (Eq.~\ref{threepoint})
can be found in Refs.~\cite{vanoldenborgh} and~\cite{veltman}. 
In certain kinematical regions the four-point function (eq.~\ref{fourpoint}) 
is also known analytically \cite{vanoldenborgh,veltman}.

\section{RPA propagators}
\label{correlators}
Using the definitions given in the previous section the gap equation
(Eq.~(\ref{gap})) takes the form
\beq
    m \= m_0 + 2ig_s\ 4 N_c N_f\ m\ I_1 \;.
\label{gapex}
\eeq
Similarly one can evaluate the quark-antiquark polarization diagrams 
(Eq.~(\ref{pol0})) and calculate the RPA meson propagators.
The results read: 
\bea
  D_\sigma(p) &=& \frac{-2 g_s}{{m_0\over m} +2i g_s\ 2 N_c N_f\ 
 (p^2-4m^2)\ I(p)}~,\label{sigma}\\   
  D_\pi(p) &=& \frac{-2 g_s}{{m_0\over m} +2i g_s\ 2 N_c N_f\ p^2\ I(p)}
 \label{pseudopi}~,\\   
  D_\rho(p) &=& \frac{-2 g_v}{1+ 2i g_v\ {4\over3} N_c N_f\ 
 (-2 m^2\ I(0)+(p^2+2 m^2)\ I(p))}\label{rho}~,\\   
  D_{a_1}(p) &=& \frac{-2  g_v}{1+ 2i g_v\ {4\over3} N_c N_f\ (-2 m^2\
    I(0)+(p^2-4 m^2)\ I(p))}~.\label{a1}
\eea   
We should comment on the $\rho$- and the $a_1$-propagator. 
A straight-forward evaluation of the vector polarization diagrams gives
\beq
    \Pi_\rho(p) =  -i {4\over3} N_c N_f\ (-2 I_1+(p^2+2 m^2)\ I(p))~.
\label{pirhoex}
\eeq
Because of vector current conservation this function should vanish
for $p^2$~=~0. This is only true if 
\beq
     m^2\,I(0) \= I_1~,
\label{mi0}
\eeq
which is not the case if we regularize $I(p)$ and $I_1$ as described
in App.~\ref{integrals}. This corresponds to the standard form of
Pauli-Villars regularization in the NJL model \cite{klevansky}. 
Alternatively one could perform the replacement Eq.~(\ref{pv}) for 
the entire polarization loop. In fact, this is more in the original
sense of Pauli-Villars regularization \cite{itzykson}. Then the factor 
$m^2$ in Eq.~(\ref{pirhoex}) should be replaced by a factor $\mu_j^2$ 
inside the sum over regulators and one can easily show that Eq.~(\ref{mi0}) 
holds (see~Eqs.~(\ref{i1}) and~(\ref{i0})). 
However, this scheme would lead to even more severe problems:
From the gap equation (Eq.~\ref{gapex}) we conclude that $i I_1$ should 
be positive. On the other hand the pion decay constant in the chiral limit
and in leading order in $1/N_c$ is given by \cite{klevansky}
\beq
f_{\pi}^{2(0)} \=  -2i N_c N_f\ m^2\ I(0)~.
\label{fpiex}
\eeq
which implies that $i m^2 I(0)$ should be negative.
So irrespective of the regularization scheme Eq.~(\ref{mi0}) cannot be 
fulfilled if we want to get reasonable results for $m$ and $f_\pi^{(0)}$
at the same time. 
Therefore we choose the standard form of Pauli-Villars regularization in the 
NJL-model \cite{klevansky} and replace the term $I_1$ in 
Eq.~(\ref{pirhoex}) by hand by $m^2\ I(0)$. This leads to the $\rho$-meson 
propagator as given in Eq.~(\ref{rho}). For consistency the $a_1$ has
been treated in the analogous way.

\section{Explicit expressions for the meson-meson vertices}
\label{functions}
In this section we list the explicit formulae for the meson-meson 
vertices. We restrict ourselves to those combinations which are needed
for the calculations presented in this article.

We begin with the three-meson vertices $\Gamma_{M_1,M_2,M_3}(q,p)$
(see~Fig.~\ref{fig4}): 
\bea
-i\Gamma_{\sigma,\sigma,\sigma}(q,p) &=& i 2 m N \Big(I(p^{\prime})
          +I(q)+I(p)+(4m^2-\frac{1}{2}(p^{\prime2}+p^2+q^2)) 
          M (p,-q)\Big) ~,
\nonumber\\
-i\Gamma^{ab}_{\pi,\pi,\sigma}(q,p) &=& i 2 m N\delta_{ab} 
\Big(I(p^{\prime})+p\!\cdot\!q M(p,-q)\Big) ~,
\nonumber\\
-i\Gamma^{\mu\lambda,ab}_{\rho,\rho,\sigma}(q,p)
&=&\delta_{ab}h(q,p)\Big(g^{\mu\lambda} 
- \frac{p^2\,q^{\mu}q^{\lambda} \+ q^2\,p^{\mu}p^{\lambda}
        \;-\;p\!\cdot\!q\,(p^{\mu}q^{\lambda}+q^{\mu}p^{\lambda})}
        {p^2q^2 \;-\; (p\!\cdot\!q)^2} \Big)~,
\nonumber\\ 
h(q,p) &=& i m N\Big(I(q)+I(p)-2 I(p^{\prime})
+(4 m^2-2 p\!\cdot\!q-p^{\prime2}) M(p,-q)\Big)~,
\nonumber\\
-i\Gamma_{\pi,\pi,\rho}^{\mu,abc}(q,p) &=&  \eps_{abc} \Big(q^{\mu}
f(q,p)-p^{\mu} f(p,q)\Big)~, \nonumber\\
f(q,p) &=& N \Big(-I(q)+p^2 M(p,-q)+2 p\!\cdot\!q M_1(q,-p)\Big)~,
\eea
with $p^{\prime}= -p-q$ and $N=4 N_c N_f$.

For the four-meson vertices we only need to consider the special cases 
needed for the diagrams (b) and (c) in Fig.~\ref{fig3}:
\bea
-i\Gamma_{\sigma,\sigma,\sigma,\sigma}(q,p,-q)&=& -N \Big\{
\frac{I(p-q)+I(p+q)}{2} + 4 m^2\ (M(p,q)+M(p,-q))
\nonumber\\ && \hspace{1cm} 
+ 2 \big(m^2\ (4 m^2-p^2-q^2)-\frac{p^2q^2}{4}\big)\
L(p,-q,p-q)\Big\}
\nonumber\\
-i\Gamma_{\sigma,\sigma,\sigma,\sigma}(q,p,-p) &=& -N \Big\{ I(p+q) + I(0)
+ 4m^2 \big( K(p)+K(q)+2 M(p,-q)\big) 
\nonumber\\ && \hspace{1cm} 
+2 p\!\cdot\!q M(p,-q) -q^2 K(q)-p^2 K(p) 
\nonumber\\ && \hspace{1cm} 
+ m^2\big( 16 m^2-4 p^2-4 q^2 + \frac{p^2 q^2}{m^2}\big) L(p,-q,0)\Big\}
\nonumber\\
-i \Gamma_{\sigma,\pi,\sigma,\pi}^{ab}(q,p,-q) 
&=&\delta_{ab} N\Big\{I(p+q)+I(p-q)+ p^2 (4m^2-q^2) L(p,-q,p-q)\Big\}
\nonumber\\
-i \Gamma_{\sigma,\pi,\pi,\sigma}^{ab}(q,p,-p) 
&=&\delta_{ab} N\Big\{-I(p+q)-I(0)- (4m^2-q^2) (K(q)-p^2 L(p,-q,0))
\nonumber\\ && \hspace{1cm}
+p^2 K(p)-2 p\!\cdot\!q\  M(p,-q)\Big\}
\nonumber\\
-i\Gamma_{\pi,\pi,\pi,\pi}^{abcd}(q,p,-q) 
&=& -N\kappa_{abcd}\Big\{I(p+q)+I(p-q)-p^2q^2 L(p,-q,p-q)\Big\}
\nonumber\\
-i\Gamma_{\pi,\pi,\pi,\pi}^{abcd}(q,p,-p) 
&=& -N\kappa_{abcd}\Big\{I(p+q)+I(0)-p^2 K(p)
\nonumber\\ && \hspace{1.7cm}
-q^2 K(q)+2 p\!\cdot\!q M(p,-q)+p^2 q^2 L(p,-q,0)\Big\}
\nonumber\\
-i\Gamma_{\rho,\sigma,\rho,\sigma}^{ab}(q,p,-q) 
&=& -2\delta_{ab}N\Big\{I(p+q)+I(p-q)+2 I(q) -p\!\cdot\!q (M(p,-q)-M(p,q))
\nonumber\\ && \hspace{1.5cm} 
+(4 m^2-2 p^2)(M(p,q)+M(p,-q))
\nonumber\\ && \hspace{1.5cm}
+m^2 (8 m^2-6 p^2+4 q^2+\frac{p^4-(p\!\cdot\!q)^2}{m^2}) L(p,-q,p-q)\Big\}
\nonumber\\
-i\Gamma_{\rho,\sigma,\sigma,\rho}^{ab}(q,p,-q) 
&=& -2\delta_{ab}N\Big\{-I(p+q)-I(0)+(p^2-4 m^2) K(p)
\nonumber\\ &&\hspace{1.5cm}
+(q^2+2 m^2) K(q) +(4 m^2-2 p\!\cdot\!q) M(p,-q)
\nonumber\\ &&\hspace{1.5cm}
+m^2 (8 m^2-2 p^2+4 q^2-\frac{p^2 q^2}{m^2}) L(p,-q,0)\Big\}
\nonumber\\ 
-i\Gamma_{\rho,\pi,\rho,\pi}^{abcd}(q,p,-q) 
&=& 2N\kappa_{abcd}\Big\{-I(p+q)-I(p-q)-2 I(q)
\nonumber\\ &&\hspace{1.5cm}
+2 p^2 (M(p,q)+M(p,-q)) +p\!\cdot\!q(M(p,-q)-M(p,q)) 
\nonumber\\ &&\hspace{1.5cm}
+(2 m^2 p^2- p^4+(p\!\cdot\!q)^2) L(p,-q,p-q)\Big\}
\nonumber\\
-i\Gamma_{\rho,\pi,\pi,\rho}^{abcd}(q,p,-q) 
&=& 2 N\kappa_{abcd}\Big\{I(p+q)+I(0)-p^2 K(p)-(q^2+2 m^2) K(q)
\nonumber\\ &&\hspace{1.5cm}
+2p\!\cdot\!q M(p,-q)+p^2 (2 m^2+q^2) L(p,-q,0)\Big\}~,
\eea
with $\Gamma_{\rho,M,M,\rho}(q,p,-q) =
g_{\mu\nu}\Gamma_{\rho,M,M,\rho}^{\mu\nu}(q,p,-q),
\Gamma_{\rho,M,\rho,M}(q,p,-p) = g_{\mu\nu}\Gamma_{\rho,M,\rho,M}^{\mu\nu}(q,p,-p)$ and $\kappa_{abcd} = \delta_{ab}\delta_{cd}+\delta_{ad}\delta_{bc}-\delta_{ac}\delta_{bd}$.
\end{appendix}

%
%
%

\end{document}